\documentclass[twocolumn,showpacs,preprintnumbers,amsmath,amssymb]{revtex4}

\usepackage{graphicx}
\usepackage{dcolumn}
\usepackage{bm}
\usepackage{amsfonts}

\newcommand{\ket}[1]{\mbox{$\mid \! #1 \, \rangle$}}
\newcommand{\bra}[1]{\mbox{$\langle \, #1 \! \mid$}}

\newcommand{\eea}{\end{eqnarray}}
\newcommand{\bea}{\begin{eqnarray}}
\newcommand{\eeas}{\end{eqnarray*}}
\newcommand{\beas}{\begin{eqnarray*}}

\newcommand{\VV}{\ensuremath{\mathcal{V}}}

\newcommand{\sz}{\sigma_z}

\newcommand{\eins}{\mbox{$1 \hspace{-1.0mm}  {\bf l}$}}

\begin{document}


\title{Linear Optics C-Phase gate made simple}

\author{Nikolai Kiesel$^{1,2}$, Christian Schmid$^{1,2}$, Ulrich Weber$^{1,2}$, Rupert Ursin$^{3}$, Harald Weinfurter$^{1,2}$}

\affiliation{
$^{1}$Max-Planck-Institut f{\"u}r Quantenoptik, D-85748 Garching, Germany\\
$^{2}$Department f\"ur Physik, Ludwig-Maximilians-Universit{\"a}t,
D-80797
M{\"u}nchen, Germany\\
$^{3}$Institut f{\"u}r Experimentalphysik,
Universit{\"a}t Wien, A-1090 Wien, Austria}%

\date{\today}

\begin{abstract}

Linear optics quantum logic gates are the best tool to generate
multi-photon entanglement. Simplifying a recent approach
\cite{Ralph_PG} we were able to implement the conditional phase
gate with only one second order interference at a polarization
dependent beam splitter, thereby significantly increasing its
stability. The improved quality of the gate is evaluated by
analysing its entangling capability and by performing full process
tomography. The achieved results ensure that this device is well
suited for implementation in various multi photon quantum
information protocols.

\end{abstract}

\maketitle

The quantum computer is one of the most promising and desirable
goals in quantum information science. Its implementation relies
strongly on the capability to engineer entanglement in the quantum
system of choice. For qubits it was shown that entangling gates
(like the C-phase gate or the CNOT) together with single qubit
operations are sufficient to create any kind of quantum network.

Photons are well suited for quantum information tasks, as their
interaction with the environment is small guaranteeing low
decoherence. While the {\em creation} of entangled photon pairs
via spontaneous parametric down conversion became a standard
technique, its {\em control} is still a major challenge, mainly
due to low nonlinear interaction efficiencies. One solution to
this problem is using linear optics components and introducing the
nonlinearity via ancillary single photons and photon number
resolving detectors \cite{KLM}. Initial demonstrations showed that
such gates can be implemented, once the necessary sources and
detectors become available on a larger scale \cite{KLM-exp}.
Another solution, requiring much less resources becomes possible
if one focuses on performing only a limited number of quantum
logic operations.  Then one can control the action of the gate by
conditioning it to the detection of one photon in each of the
output ports. This will occur only with a certain probability,
which, however, might be larger than the one achieveable with the
first method and equivalent resources. In particular, a controlled
phase (C-Phase) gate was introduced recently \cite{Ralph_PG},
which uses a combination of first and second order interference to
obtain C-Phase action in 1/9 of the cases. Yet, since first-order
interference requires stability of the setup on the order of less
than the photon's wavelength, for multi-photon experiments
\cite{schmid} more simple and stable implementations surely are
desirable.

Here we introduce a linear optics C-Phase gate, which uses only a
single two-photon interference at a polarization dependent beam
splitter. The stability requirements are thereby relaxed to the
coherence length of the detected photons ($\approx150 \mu m$) and
can easily be fulfilled without additional stabilization
equipment. To characterize the C-phase gate we first study the
entangling capability of the gate by determining the fidelity and
negativity of the output for four different input states. Second,
we use linear quantum process tomography (QPT)
\cite{linQPT,Steinberg} to analyze the gate operation. As
imperfect interference reduces the quality of the gate and induces
state-selective incoherence we had to account for the
non-trace-preserving character of the gate. Instead of the usual
maximum likelihood approach, we use prior knowledge of the
intrinsic features of our setup, in order to obtain physical and
easily understandable parameters for characterizing the gate and
estimating its performance.

\begin{figure}
\includegraphics[width=8.5cm,clip]{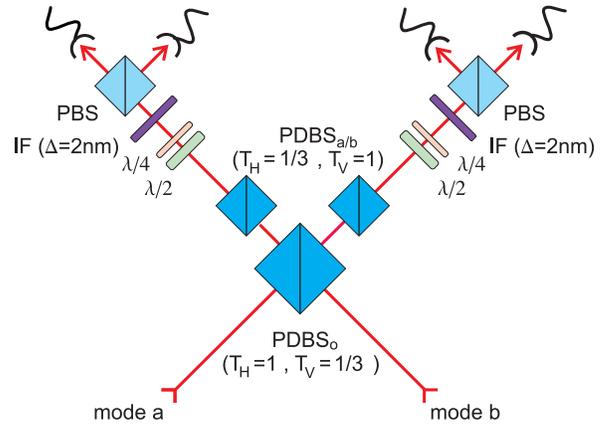}
\caption{Experimental setup for the C-Phase gate. The phase is
introduced by a second order interference at a polarization
dependent beam splitter $\mathrm{PDBS_O}$. To obtain equal output
amplitudes for any input polarization state polarization dependent
beam splitters with reversed splitting ratio $\mathrm{PDBS}_{a/b}$
are placed in each mode. The gate operation is applied in case of
a coincidence detection between modes a and b. The resulting
output state is analyzed via half- and quarter-wave plates
$\lambda/2$, $\lambda/4$ and a polarizing beam splitter
PBS.}\label{fig:pgsetup}
\end{figure}

The ideal C-Phase gate acts on two-qubit input states
\begin{eqnarray}
\ket{\psi_{in}}&=&(c_{HH}\ket{HH}+c_{HV}\ket{HV}\\ \notag
&+&c_{VH}\ket{VH}+c_{VV}\ket{VV})\,, \label{eq:ginput}
\end{eqnarray}
and applies a relative $\pi$-phaseshift to the contribution VV
only, such that
\begin{eqnarray}
\ket{\psi_{out}}&=&(c_{HH}\ket{HH}+c_{HV}\ket{HV}\\ \notag
&+&c_{VH}\ket{VH}-c_{VV}\ket{VV})\,. \label{eq:idoutput}
\end{eqnarray}
Here we encode the logical 0 (1) in linear horizontal $H$
(vertical $V$)
 polarization of single photons. $c_{HH}$ denotes the amplitude of the
$\ket{HH}$-term (for the other terms accordingly).

The application of the phase shift relies on second-order
interference of indistinguishable photons at a polarization
dependent beam splitter (PDBS) (Fig.\ref{fig:pgsetup})
\cite{HOM,CST}. Two input modes a and b are overlapped at
$\mathrm{PDBS_O}$. The transmission of $1/3$ for vertical
polarization results in a total amplitude of $-1/3$ for the
$\ket{VV}$ output terms, as can be seen by adding the amplitudes
for a coincident detection:
\begin{equation}
(t_V^{a} \cdot  t_V^{b})+(i r_V^{a} \cdot i
r_V^{b})=\sqrt{\frac{1}{3}}\sqrt{\frac{1}{3}}-\sqrt{\frac{2}{3}}\sqrt{\frac{2}{3}}=-1/3
\end{equation}
where $t_i^x$ ($r_i^x$) is the amplitude for transmission
(reflection) of state $\ket{i}$ in mode $x$. Perfect transmission
of horizontal polarization causes that no interference happens on
the contributions $\ket{HH},\ket{HV}$ and $\ket{VH}$. As the
absolute values of the amplitudes need to be equal for any input
we still need to attenuate the contributions that include
horizontal polarization. This is achieved by $\mathrm{PDBS_{a/b}}$
with the transmission $1/3$ for horizontal polarization and
perfect transmission for vertical polarization in both output
modes. All together we find a probability of $1/9$ to obtain a
coincidence in the outputs and thus a gate operation with perfect
fidelity.

Working with real components results in deviations from the
theoretical derivation. A detailed calculation with arbitrary
transmission and reflection amplitudes at $\mathrm{PDBS_{O}}$ and
$\mathrm{PDBS_{a,b}}$ shows how their parameters influence the
gate operation. In general we obtain from $\ket{\psi_{in}}$
\begin{equation}
\begin{array}{l}
\ket{\psi_{out}}=
(c_{HH}t_{H}^{a}t_{H}^{b}a_{H}b_{H}-c_{HH}r_{H}^{a}r_{H}^{b}a_{H}b_{H})\ket{HH}\\
+(c_{HV}t_{H}^{a}t_{V}^{b}a_{H}b_{V}-c_{VH}r_{V}^{a}r_{H}^{b}a_{H}b_{V})\ket{HV}\\
+(c_{VH}t_{V}^{a}t_{H}^{b}a_{V}b_{H}-c_{HV}r_{H}^{a}r_{V}^{b}a_{V}b_{H})\ket{VH}\\
+(c_{VV}t_{V}^{a}t_{V}^{b}a_{V}b_{V}-c_{VV}r_{V}^{a}r_{V}^{b}a_{V}b_{V})\ket{VV})\,,
\end{array}
\label{eq:realoutput}
\end{equation}\\
where $a_i$ ($b_i$) are the transmission amplitudes of $\ket{i}$
in mode a (b). To obtain the expected C-phase gate operation one
has to fulfill several conditions, which give an insight in how
the setup has to be built. First,
$(r_{V}^{a}r_{V}^{b})/(t_{V}^{a}t_{V}^{b})=2$, which is
approximately achieved by slightly varying the angle of incidence
at $\mathrm{PDBS_O}$. Experimentally we reach a value of $2.018
\pm 0.003$. Second, $r_{H}^{a}=0=r_{H}^{b}$, which requires the
reflection of the horizontal polarization at the overlap beam
splitter to be zero. The third condition,
$t_{H}^{a}a_{H}=t_{V}^{a}a_{V}$, and
$t_{H}^{b}b_{H}=t_{V}^{b}b_{V}$, respectively, determines the
setting for the attenuation at $\mathrm{PDBS_{a,b}}$.

\begin{figure}
\includegraphics[width=8.5cm,clip]{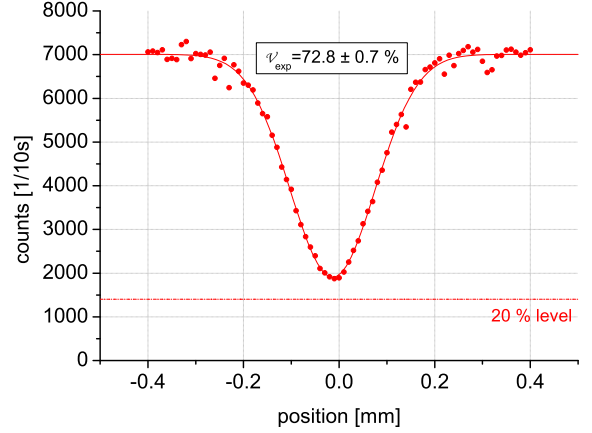}
\caption{HOM-Interference at the polarization dependent overlap
beam splitter in the phase gate for a \ket{VV} input. In case of
perfect interference the countrate should drop down to 20 \%
leading to a theoretically achievable dip visibility of 80 \% .}
\label{fig:HOM}
\end{figure}

 To experimentally test the gate operation
we used photon pairs emitted from spontaneous parametric down
conversion. A 2 mm thick BBO ($\beta$-Barium Borate) crystal was
pumped by UV pump pulses with a central wavelength of 390~nm and
an average power of 700~mW from a frequency-doubled mode-locked
Ti:sapphire laser (pulse length 130~fs). The pulsed operation is
not necessary when working only with photon pairs, but as the gate
is intended to work in multi-photon applications we preferred to
characterize it for this mode of operation. The emission is
filtered with polarizers to prepare input product states with high
quality. We couple the photon pairs into single mode fibers for
selection of the spatial modes. This guarantees identical beam
modes which eases the alignment of spatial mode matching at
$\mathrm{PDBS_O}$. The spectral mode selection is improved via 2nm
bandwidth filters behind the gate.

To ensure the same optical path length between the crystal and the
overlap beam splitter for both photons, one of the output couplers
of the single mode fibers is mounted on a translation stage. The
position of zero delay is determined from the minimum of the
coincidence rate for $\ket{VV}$-input (Hong-Ou-Mandel \cite{HOM},
"HOM", Dip Fig. \ref{fig:HOM}). In each output mode of the C-Phase
gate the polarization is analyzed via quarter and half waveplates
and a polarizing beam splitter with single photon avalanche photo
diodes. For the analysis of the final two-photon states the
coincidence count rates for each of the four contributions have to
be corrected for the different detector efficiencies. The errors
on all quantities are deduced from propagated Poissonian
counting statistics of the raw detection events and efficiencies.\\
\begin{figure}
\includegraphics[width=8.0cm,clip]{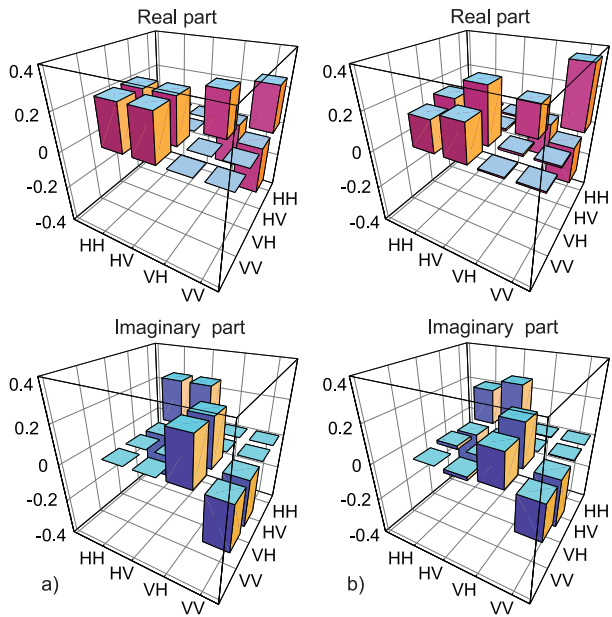}
\caption{(a) Theoretically expected and (b) experimentally
obtained gate output density matrix for \ket{L+} input state.}
\label{fig:rho}
\end{figure}
The HOM-measurement shown in Fig. \ref{fig:HOM} also gives
information about the indistinguishability of the photons at the
PDBS. For large delay, the two photons are completely
distinguishable due to their time of arrival. The probability to
get a coincidence from a $\ket{VV}$-input is then $5/9$. In case
of perfectly indistinguishable photons at zero delay, the
probability drops to $1/9$. The Dip-Visibility is defined by
$\VV=\left(c_{\infty}-c_{0}\right)/{c_{\infty}}$, where $c_{0}$ is
the count rate at zero delay and $c_{\infty}$ at positions with
big delay. From the above considerations we obtain a theoretical
value of $\VV_{th}=80\%$, and experimentally, applying
least-square fit, we find $\VV_{exp}=72.8\% \pm 0.7\%$. We call
$\mathcal{Q}=\VV_{exp}/\VV_{th}=91.0\% \pm 0.9\%$ the overlap
quality. One can conclude that the amount of additional
$\ket{VV}\bra{VV}$-noise depends on the input, but is $9\%$ at
maximum.

\begin{figure*}
\includegraphics[width=\textwidth,clip]{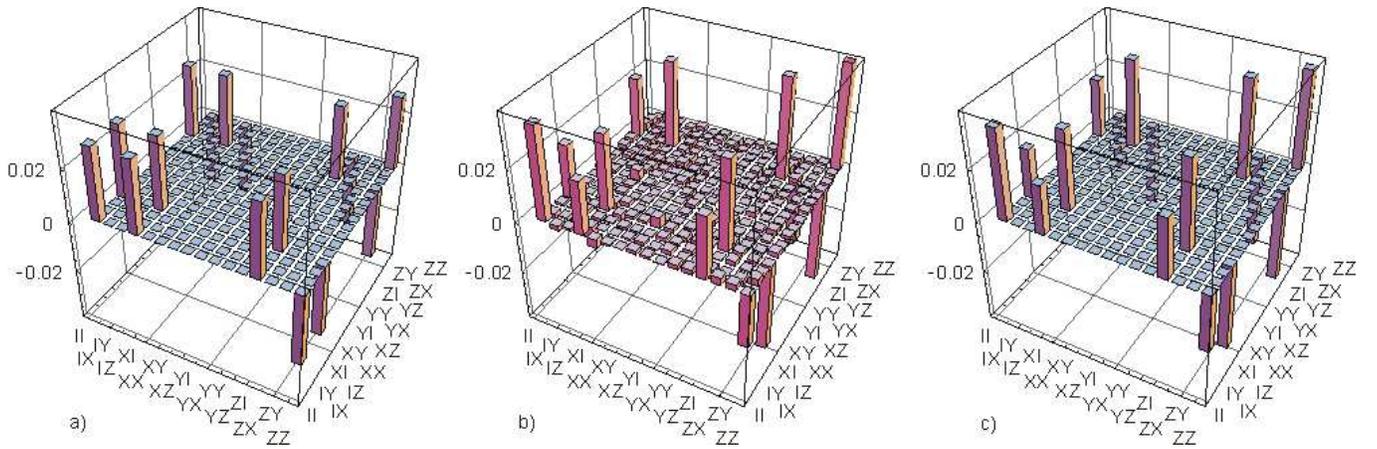}
\caption{(a) $\chi$ matrix of the QPT for an ideal phase gate, (b)
for the experimentally realized gate and (c) for a theoretical
model fit to the experimental data. The imaginary part of the
experimentally obtained $\chi$ consists of noise only which is
comparable to the one in the real part.} \label{fig:teletomo}
\end{figure*}

 As a first step in the analysis of the performance of our gate
we look at its capability to entangle. We choose
$\ket{++},\ket{+L},\ket{L+}$, and $\ket{LL}$, with
$\ket{+}=1/\sqrt{2}(\ket{H}+\ket{V})$ and
$\ket{L}=1/\sqrt{2}(\ket{H}+i\ket{V})$, as input product states
and perform state tomography on the output states \cite{MOQ}. We
use linear tomography as the resulting matrices all have
eigenvalues greater or equal -0.02, i.e., are almost physical
without corrections. For an ideal C-phase gate one would obtain a
maximally entangled output for these input states, for example
\begin{eqnarray}
\ket{L+}&=&1/2(\ket{HH}+i\ket{VH}+\ket{HV}+i\ket{VV}) \notag\\
\overline{PG}\ket{L+}&=&1/2(\ket{HH}+i\ket{VH}+\ket{HV}-i\ket{VV})\notag\\
 &=&1/\sqrt{2}(\ket{LH}+\ket{RV})\,. \label{eq:example}
\end{eqnarray}
The experimentally observed fidelities relative to the expected
output states are all better $F_{exp}\geq 80.5\% \pm 0.6 \%$. Fig.
\ref{fig:rho} exemplarily shows the experimental result for
$\ket{L+}$-input ($F_{exp}^{L+}=87.8 \% \pm 0.6 \%$). Note that
for states with a fidelity larger than $(2+3\sqrt{2})/8 = 0.78$
CHSH-inequalities are violated (\cite{CHSH}), which is the case
for all of our examples. To quantify the entanglement we also
calculated the logarithmic negativity --- for all output states we
find $\mathcal{N}_{exp} \geq 0.73 \pm 0.02$
($\mathcal{N}_{exp}^{L+} = 0.75 \pm 0.02$).

For a complete characterization of an arbitrary unknown process
one can use quantum process tomography (QPT).
For QPT the process is represented by a superoperator
$\mathcal{E}$ which is decomposed in a linear combination of a
basis of unitary transformations $E_i$:
\begin{equation}
\mathcal{E}(\rho_{in})=\sum_{i,j} \chi_{ij} E_i \rho_{in}
E_j^{\dagger}
\end{equation}
The matrix $\chi$ completely describes the process. In order to
obtain all components $\chi_{ij}$, the normalized output density
matrices $\rho_{out}^{k}$ for a tomographic set of, usually
separable, input states are measured, in our case for the inputs
($\ket{HH}$ , $\ket{HV}$ , $\ket{H+}$ , $\ket{H-}$ , $\ket{VH}$ ,
$\ket{VV}$ , $\ket{V+}$ , $\ket{VL}$ , $\ket{+H}$ , $\ket{+V}$ ,
$\ket{++}$ , $\ket{+L}$ , $\ket{LH}$ , $\ket{LV}$ , $\ket{L+}$ ,
$\ket{LL}$). As the contribution of the $\ket{VV}\bra{VV}$-noise
is input state dependent our process is non-trace-preserving. This
means that the outcomes occur with different probabilities $p_k$
\cite{linQPT} for the different input states $\rho_{in}^{k}$
$\mathrm{Tr}(\mathcal{E}(\rho_{in}^{k}))$:
\begin{multline}
\rho_{out}^{k}=\frac{\mathcal{E}(\rho_{in}^{k})}{\mathrm{Tr}(\mathcal{E}(\rho_{in}^{k}))}\\
\Rightarrow
\mathcal{E}(\rho_{in}^{k})=\mathrm{Tr}(\mathcal{E}(\rho_{in}^{k}))\rho_{out}^{k}=p_k
\rho_{out}^{k}.
\end{multline}
We determine these probabilities from the diagonal entries of all
measured output density matrices. The normalized density matrices
together with the corresponding probabilities can be used to
evaluate $\chi_{ij}$ via Eq. (7) and (6).  To account for the
probabilistic nature of the gate an overall normalization is
performed such that $p_{HH}=1/9$.

Fig. \ref{fig:teletomo}a shows the process matrix $\chi_{th}$ of
the ideal linear optics phase gate. It represents the
decomposition of the C-Phase gate into unitary operations, for our
choice of $E_i$ resulting in
\begin{equation}
\overline{PG}_{ideal}=\left(\eins\otimes\eins+\sz\otimes\eins+\eins\otimes\sz-\sz\otimes\sz\right)/3\,.
\end{equation}
The four peaks in the diagonal of $\chi_{th}$ show the equal
weights of the contributions, while the negative entries at the
edges represent the negative sign at $\sz\otimes\sz$. This matrix
now can be compared with the experimentally obtained one (Fig.
4b). Only the real parts are shown since the imaginary parts are
close to zero (average $0.0 \pm 0.002$). As the introduced noise
is not too big, the experimentally measured process matrix still
demonstrates nicely the features of the gate operation. The main
differences are the
lower non-diagonal terms indicating reduced coherence in the
system. From the estimated process tomography matrix we calculated
a process fidelity of
$F_p=Tr(\chi_{th}.\chi_{exp})/(\mathrm{Tr}(\chi_{th}).\mathrm{Tr}(\chi_{exp}))=81.8\%$.
Still, due to Poissonian counting statistics $\chi_{exp}$ has
non-physical, negative eigenvalues, and the above value has to be
treated with care. To circumvent this problem one can use the
maximum likelihood approach, where a physical process matrix is
fitted to the observed data.  Yet, the process is not really
unknown to us and we can try to describe it via a theoretical
model according to equation \ref{eq:realoutput}.

The transformation of the phase gate consists of interference
between both photons transmitted or both photons reflected
$\overline{PG}_{gen}=M_{tt}+M_{rr}$, where both $M_{tt}$ and
$M_{rr}$ are 
matrices with components given by the coefficients of Eq.
(\ref{eq:realoutput}). For simplicity we assume $t_V^a=t_V^b$ and
$r_H^a = r_H^b =0$. $M_{rr}=|r_V|^2\ket{V}\bra{V}$ reduces then to
only one nonvanishing matrix element. The state dependent noise
originates from the fact that interference occurs only with a
probability according to the quality parameter $\mathcal{Q'}$ and
is incoherent otherwise, which finally yields

\begin{eqnarray}
\overline{PG}_{mod}\rho\overline{PG}_{mod}^\dagger&=&\mathcal{Q'}(M_{tt}+M_{rr})\rho(M_{tt}+M_{rr})^\dagger\\
\notag
 &+&(1-\mathcal{Q'})(M_{tt} \rho M_{tt}^\dagger + M_{rr}\rho
M_{rr}^\dagger)
\end{eqnarray}


From this ansatz we obtain a model QPT-matrix $\chi_{mod}$ by
minimizing the sum of the absolute squared values of all the
matrix elements of $\chi_{mod}-\chi_{exp}$ numerically (see Fig.
\ref{fig:teletomo}c). The obtained quality value
$\mathcal{Q'}=0.904$ is in very good agreement with $\mathcal{Q}$
obtained from the fit to the HOM-dip. This indicates that it is
indeed mainly imperfect overlap at the beam splitter which causes
the state dependent noise. In order to compare the model with the
real setup we calculate the fidelities between the predicted and
the experimentally measured output density matrices obtaining an
average value of F$^{exp}_{mod \mathcal{Q}'}=96.6\% \pm 1.7\%$.
An alternative model including depolarization in the gate did not
significantly change the figure, the resulting white noise was
negligible.

 In conclusion we have presented a C-Phase gate acting on the polarization degree of photons. The
gate relies only on one second order interference at a
polarization dependent beam splitter and thus significantly
simplifies previous approaches. We have demonstrated the
entangling quality of the gate for various input states. A linear
quantum process tomography allowed us to match a model of the gate
to the experimental data. The resulting fit proofs the assumption,
that the main deviation from optimal performance is due to
distinguishability of incident photons at the overlap beam
splitter. By means of further filtering this can be improved on
the cost of count rate. The results ensure that this gate is ready
to be used in various quantum information processing tasks such as
generating multi photon entanglement or for complete Bell state
analysis in quantum teleportation and entanglement swapping
experiments.

We acknowledge stimulating discussions with A. Zeilinger. This
work was supported by the Deutsche Forschungsgemeinschaft and the
European Commission through the EU Project RamboQ (IST-2001-38864)


\begin{thebibliography}{99}

\bibitem{Ralph_PG} T.C.~Ralph, N.K.~Langford, T.B.~Bell, and A.G.~White, Phys. Rev. A {\bf 65}, 062324
(2002); H.F.~Hofmann and S.~Takeuchi, Phys. Rev. A {\bf 66},
024308 (2002); J.L.~O'Brien, G.J.~Pryde, A.G.~White, T.C.~Ralph
and D.~Branning, Nature {\bf426}, 264 (2003); J.L.~O'Brien et al.,
Phys. Rev. Lett. {\bf 93}, 080502 (2004).

\bibitem{KLM}E.~Knill, R.~Laflamme, G.J.~Milburn, Nature 409, 46-52 (2001).

\bibitem{KLM-exp}T.B.~Pittman, M.J.~Fitch, B.C.~Jacobs, and J.D.~Franson, Phys. Rev. A {\bf68}, 032316 (2003);
K.~Sanaka, T.~Jennewein, J.W.~Pan, K.~Resch, A.~Zeilinger, Phys.
Rev. Lett. 92, 017902 (2004); S.~Gasparoni, J.W.~Pan, P.~Walther,
T.~Rudolph, A.~Zeilinger, Phys. Rev. Lett. {\bf 93}, 020504
(2004).



\bibitem{schmid} C.~Schmid, et al., to be published; N.~Kiesel, et al. to be
published.





\bibitem{linQPT} I.L.~Chuang, and M.A.Nielsen, J. Mod. Opt.
{\bf44}, 2455 (1997); J.F.~Poyatos, and  J.I.~Cirac, P.~Zoller,
Phys. Rev. Lett. {\bf 78}, 390 (1997).

\bibitem{Steinberg} M.W.~Mitchell, C.W.~Ellenor, S.~Schneider, and A.M.~Steinberg, Phys. Rev. Lett. {\bf91},
120402 (2003).

\bibitem{HOM}C.K.~Hong, Z.Y.~Ou, L.~Mandel, Phys. Rev. Lett. {\bf 59}, 2044
(1987).

\bibitem{CST}R.A.~Campos, B.E.A.~Saleh, and M.C.~Teich,
Phys. Rev. A {\bf 42}, 4127 (1990).

\bibitem{MOQ}  D.F.V.~James, P.G.~Kwiat, W.J.~Munro and A.G.~White, Phys. Rev. A {\bf 64}, 052312 (2001).

\bibitem{CHSH} C.H.~Bennett, G.~Brassard, S.~Popescu, B.~Schumacher, J.A.~Smolin, and W.K.~Wootters, Phys. Rev. Lett. {\bf 76}, 722 (1996).

\bibitem{fitpar}Other parameters are determined as:
$t_V^2/r_V^2=2.035$, $a_Ht_H=1.0a_Vt_V$, $b_Ht_H=1.16b_Vt_V$














\end{thebibliography}
\end{document}